\documentclass[11pt]{article}
\usepackage{times,graphicx,theorem,epsfig,lscape,amsmath,dcolumn,multirow,epic,float,enumerate,comment}
\usepackage{natbib}
\usepackage{amsmath,amssymb}
\usepackage[normalem]{ulem}
\usepackage{subfigure}
\usepackage{natbib}
\usepackage[format=hang,labelfont=bf]{caption}
\usepackage[figuresright]{rotating}
\usepackage[english]{babel}
\usepackage{booktabs}
\usepackage{hhline}
\usepackage{bm}
\usepackage{enumerate}
\usepackage{verbatim}
\usepackage{dsfont}
\usepackage{threeparttable}
\usepackage{array}
\usepackage{url}

\newcommand{\blinding}[2]{#1}   %%% USE THIS FOR UNBLINDED
\newcommand{\bfX}{\mathbf{X}}

\newcommand{\E}{\mathbb{E}}
\newcommand{\CFD}{\mbox{\tiny{CFD}}}
\newcommand{\CMF}{\mbox{\tiny{CMF}}}
\newcommand{\NB}{\mathrm{NB}}
\newcommand{\reg}{\mathrm{reg}}

\newcommand{\wt}{\mathrm{wt}}

\newcommand{\dr}{\mathrm{dr}}

\newcommand{\bfbeta}{\bm{\beta}}
\newcommand{\bfgamma}{\bm{\gamma}}
\newcommand{\bfalpha}{\bm{\alpha}}

\newtheorem{assumption}{{\sc Assumption}}

\newcommand{\sumi}{\sum_{i=1}^N}
\newcommand{\cp}{\stackrel{p}{\rightarrow}}

\theoremstyle{plain} 
\theoremstyle{plain} \newtheorem{proposition}{{\sc Proposition}}
\theoremstyle{plain} 
\theoremstyle{plain}

\begin{document}

\begin{center}
\vspace*{-2.5cm}

{\Large Double-Robust Estimation in Difference-in-Differences with an Application to Traffic Safety Evaluation}

\medskip
\blinding{
Fan (Frank) Li\footnote{Fan (Frank) Li is PhD student, Department of Biostatistics and Bioinformatics, Duke University, Durham, NC, 27710 (email: frank.li@duke.edu); $^2$Fan Li is Associate Professor, Department of Statistical Science, Duke University, Durham, NC, 27705 (email: fli@stat.duke.edu).} \quad and \quad Fan Li$^2$

%\today
}{}

\end{center}

\date{}

{\centerline{ABSTRACT}

\noindent Difference-in-differences (DID) is a widely used approach for drawing causal inference from observational panel data. Two common estimation strategies for DID are outcome regression and propensity score weighting. In this paper, motivated by a real application in traffic safety research, we propose a new double-robust DID estimator that hybridizes regression and propensity score weighting. We particularly focus on the case of discrete outcomes. We show that the proposed double-robust estimator possesses the desirable large-sample robustness property. We conduct a simulation study to examine its finite-sample performance and compare with alternative methods. Our empirical results from a Pennsylvania Department of Transportation data suggest that rumble strips are marginally effective in reducing vehicle crashes.
\vspace*{0.3cm}

}

\clearpage

\section{Introduction}
\label{sec:intro}
Difference-in-differences (DID) is a popular evaluation strategy used across a range of disciplines. It uses data with a time dimension to control for unobserved but fixed confounding, and identifies causal effects by contrasting the change in outcomes pre- and post-treatment, among the treated and control groups \citep{Ashenfelter1978,Ashenfelter1985}.  The most common DID setting is a before-after design, in which the treated and control units are genuinely comparable. For example, DID often exploits a policy shift that occurred in one region but not in an adjacent region \citep{Card1994}. The key assumption of DID is \emph{parallel trend}, that is, the counterfactual trend behavior of treatment and control groups, in the absence of treatment, is the same \citep{Heckman1997}.

The target causal estimand in DID is a version of the average treatment effect for the treated (ATT). Estimation of ATT in DID is traditionally tied with a fixed-effects outcome regression model \citep{Angrist2009}. Though flexible, the regression method relies on strong assumptions such as homogenous and additive effects, and can be sensitive to model misspecification. Alternatively, \cite{Abadie2005} proposed a semiparametric estimator for DID based on inverse probability weighting (IPW) where only a model for the propensity score but not the outcome is required. The IPW estimator does not require assumptions on the outcome distribution, but may be inefficient compared to a correctly-specified outcome model. Outside the DID context, the double-robust (DR) method \citep{Bang2005} that augments an IPW estimator by an outcome regression have received much attention in causal inference. A DR estimator is consistent if either the outcome model or the propensity score model, but not necessarily both, is correctly specified \citep{Scharfstein1999}, and it is semiparametrically efficient when both models are correctly specified \citep{Robins1994,Robins1995analysis,Robins1997towards}. However, most DR methods focus on the average treatment effect (ATE) estimand rather than ATT. In this paper, the intrinsic connection between Abadie's DID estimator and the IPW technique motivates us to devise a new double-robust DID estimator for ATT, and we show it possesses the desirable large-sample robustness property.

Our method is motivated from a real application to traffic safety research. Specifically, we wish to evaluate the impact of installing rumble strips---a low-cost traffic safety countermeasure---on vehicle crashes. Due to ethical and practical constraints with roadway safety experimentation, observational studies are routinely used for such evaluations. The state-of-the-art method in traffic safety evaluation---the Empirical Bayes (EB) approach---adopts a treatment-control before-after design \citep{Hauer1997}, where the crash outcomes in a number of comparable treated and control sites were recorded both before and after the safety countermeasure was installed. This design fits naturally into the DID framework, but to our knowledge, the connection was never made in the literature. In fact, the EB approach is entirely regression-based and comes without a causal interpretation. As a robust alternative to EB, a recent stream of research advocated propensity score methods to the after period data alone \citep{Karwa2011, Wood2015a,Wood2015b, Wood2016a}. However, ignoring the data in the before period may present a major information loss and fail to adjust for the time trend. In contrast, our proposed double-robust estimator combines the virtues of the regression-based and the propensity score weighting estimators for before-after studies. Because the outcome is count data in the transportation application, we particularly focus on the case of discrete outcomes in our modeling and estimation.

The rest of the paper is organized as follows. Section \ref{sec:did} defines the causal estimands, and introduces DID estimators: outcome regression, propensity score weighting and the proposed double-robust estimators. Section \ref{sec:app} presents the application to highway crash data collected by the Pennsylvania Department of Transportation. Section \ref{sec:sim} further illustrates the DID estimators through simulations mimicking the traffic safety study. Section \ref{sec:dis} concludes.

\section{Causal Inference via Difference-in-Differences}
\label{sec:did}
\subsection{Causal Estimands}
We introduce the notation in the context of the evaluation of rumble strips (i.e., treatment). We consider the basic two-period two-group DID design. Assume a sample of traffic sites---units of analysis---indexed by $i=1,\ldots,N$, belong to one of the two groups, with $G_i=1$ indicating that rumble strips were applied in the after period, i.e. the treatment group, and $G_i=0$ indicating that rumble strips were not applied in either period, i.e., the control group. Units in both groups are followed in two periods of time, with $T=t$ and $T=t+1$ denoting the before and after period, respectively. For each unit $i$, let $D_{iT}$ be the observed treatment status at period $T$. Since none of the traffic sites received treatment in the before period, we have $D_{it}=0$ for all $i$. Because the treatment is only administered to one group ($G_i=1$) in the after period, $D_{i,t+1}=1$ for all units in group $G_i=1$ and $G_i=D_{i,t+1}$ for all $i$. Similar to prior traffic safety evaluation studies \citep{Karwa2011,Wood2017}, we make the Stable Unit Treatment Value Assumption (SUTVA), meaning no interference between units and no different versions of the treatment. This assumption is more reasonable when the traffic sites are far apart from one another, but may be questionable when the sites are in close proximity. We will proceed with this assumption and discuss in Section \ref{sec:dis} the implications when SUTVA is violated. Under SUTVA, each unit has two potential crash counts in each period, $Y_{iT}(0)$ and $Y_{iT}(1)$, and only the one corresponding to the observed treatment status, $Y_{iT}=Y_{iT}(D_{iT})$, is observed. The DID design allows us to write $Y_{it}=Y_{it}(0)$ and $Y_{i,t+1}=(1-G_i)Y_{i,t+1}(0)+G_iY_{i,t+1}(1)$. A vector of $p$ pre-treatment variables, $\bfX_i$, is also observed for each unit. We denote the collection of observed data by $\mathcal{Z}_i=\{Y_{it},Y_{i,t+1},G_i,\bfX_i\}$, and assume that the $\mathcal{Z}_i$'s are independent and identically distributed from some common distribution $\mathbb{F}(\mathcal{Z})$.

In traffic safety studies, safety countermeasures are usually applied only to selected pilot sites before rolling out to a larger scale. The safety effectiveness is usually evaluated in a multiplicative fashion using the Crash Modification Factor \citep*[CMF,][]{AASHTO2010}, which can then be used to understand the expected change in crash frequency after a traffic safety countermeasure is implemented. In our traffic application, the rumble strip installation is implemented as part of a national safety improvement program, and the interest lies in quantifying its potential effectiveness among the sites where the rumble strips were installed. Similar to  \citet{Wood2017}, we formally define the CMF as a causal estimand that characterizes the ratio between the expected observed outcome after the installation and the expected counterfactual outcome had the countermeasure not been installed in the pilot sites. Using the potential outcome notation, we define the CMF
\begin{equation}\label{eq:cmf}
\tau_{\CMF}\equiv \frac{\E[Y_{i,t+1}(1)|G_i=1]}{\E[Y_{i,t+1}(0)|G_i=1]}=\theta_1/\theta_0,
\end{equation}
where we denote $\theta_1=\E[Y_{i,t+1}(1)|G_i=1]$ and $\theta_0=\E[Y_{i,t+1}(1)|G_i=0]$. Because the crash outcomes are count data, $\tau_{\CMF}$ is a causal rate ratio that quantifies the relative average change in crash counts due to rumble strip installation among the treated. The scale-free $\tau_{\CMF}$ is a ratio version of the usual average treatment effect among the treated (ATT) estimand. Here, to characterize the causal rate difference in the absolute scale, we also define an additive version---the Crash Frequency Difference (CFD):
\begin{equation}\label{eq:cfd}
\tau_{\CFD} \equiv \E[Y_{i,t+1}(1)-Y_{i,t+1}(0)|G_i=1] = \theta_1-\theta_0.
\end{equation}
We argue that using the pair of parameters $(\tau_{\CFD},\tau_{\CMF})$ instead of $\tau_{\CMF}$ alone presents a more complete picture of the effectiveness of safety countermeasure.

\subsection{Assumptions}\label{sec:aspt}
Estimands $\tau_{\CFD}$ and $\tau_{\CMF}$ are functions of $\theta_1$ and $\theta_0$. Under SUTVA, $\theta_1$ is nonparametrically identified: $\theta_1=\E[Y_{i,t+1}|G_i=1]$, with a consistent moment estimator
\begin{equation}\label{eq:mom}
\hat{\theta}_1={\sumi G_iY_{i,t+1}}\Big/{\sumi G_i}.
\end{equation}
In contrast, $\theta_0$---the expected counterfactual outcome in the absence of treatment at time $T=t+1$---must rely on additional restrictions to identify. Following the convention in DID design, we impose the parallel trend assumption,

\begin{assumption}\label{asp:pt}(Parallel Trend) For each unit $i=1,\ldots,N$,
$$\E[Y_ {i,t+1}(0)-Y_{it}(0)|\bfX_i,G_i=1]=\E[Y_{i,t+1}(0)-Y_{it}(0)|\bfX_i,G_i=0].$$
\end{assumption}

Assumption \ref{asp:pt} imposes that, conditional on the pre-treatment covariates $\bfX_i$, the average outcomes in the treated and control groups, in the absence of treatment, would have followed a parallel path over time. The quantity $\theta_0$ is therefore identified under Assumption \ref{asp:pt} as
\begin{equation}\label{eq:theta0}
\begin{split}
\theta_0=&\E_{\bfX}\{\E[Y_{i,t+1}(0)|\bfX_i,G_i=1]|G_i=1\}\\
=&\E_{\bfX}\{\E[Y_{it}(0)|\bfX_i,G_i=1]+\E[Y_{i,t+1}(0)-Y_{it}(0)|\bfX_i,G_i=0]|G_i=1\}\\
=&\E[Y_{it}|G_i=1]+\E_{\bfX}\{\E[Y_{i,t+1}-Y_{it}|\bfX_i,G_i=0]|G_i=1\},
\end{split}
\end{equation}
where both terms of the right hand side of the equation involve only expectations of observed data and are identified.

It is important to note that a direct DID estimator that uses \begin{equation}\label{eq:direct}
   \hat{\theta}_{0}^\text{direct}=\frac{\sumi G_iY_{it}}{\sumi G_i}+\frac{\sumi (1-G_i)(Y_{i,t+1}-Y_{it})}{\sumi (1-G_i)}.
    \end{equation}
to estimate $\theta_0$ neglects the pre-treatment covariate information and is subject to selection bias. In fact, $\hat{\theta}_{0}^\text{direct}$ is only consistent to $\theta_0$ under the unconditional version of the parallel trend assumption, i.e., $\E[Y_ {i,t+1}(0)-Y_{it}(0)|G_i=1]=\E[Y_{i,t+1}(0)-Y_{it}(0)|G_i=0]$, which is arguably stronger than Assumption \ref{asp:pt}. On the other hand, unlike the standard unconfoundedness condition usually assumed for the cross-sectional data, Assumption \ref{asp:pt} does not necessarily assume that $\bfX$ controls for all sources of confounding. Indeed, DID allows for unobserved confounders to affect treatment assignment as long as their impact on the potential outcomes is both separable and time-invariant \citep{Lechner2011}. Assumption \ref{asp:pt} is generally untestable and may be questionable in practice. As an indirect way to assess the plausibility of parallel trend, in this application, we will conduct a ``no treatment'' evaluation by performing DID analyses for crash outcomes from two pre-treatment periods ($T=t-1$ and $T=t$). Specifically, if the parallel trend assumption is plausible, that is,
\begin{equation*}
\E[Y_ {it}(0)-Y_{i,t-1}(0)|\bfX_i,G_i=1]=\E[Y_{it}(0)-Y_{i,t-1}(0)|\bfX_i,G_i=0],
\end{equation*}
then the estimated CFD and CMF based on time $T=t-1,t$ should be close to $0$ and $1$, respectively, because in reality rumble strips were not applied until after time $t$ and should have no causal effect for the pre-treatment outcomes. This idea is similar to the falsification endpoints or negative control idea in assessing unconfoundedness \citep{Rosenbaum2002}.

As in most ATT estimation, we also assume \emph{weak overlap}, that is, each unit has a nonzero probability of receiving the control, $e(\bfX_i)\equiv\Pr(G_i=1|\bfX_i)<1$, where $e(\bfX_i)$ is the propensity score. The weak overlap assumption is directly testable by visually comparing the estimated propensity score distributions between the treatment groups.

\subsection{Extant Methods: Regression and Weighting}\label{sec:exist}
Two main classes of existing estimating methods of DID are outcome regression and propensity score weighting. We first introduce a regression-based estimator specifically for count outcomes. To identify $\theta_0$, we need to identify all components on the right hand side of equation (\ref{eq:theta0}). Similar to $\hat{\theta}_1$, the first term $\E[Y_{it}|G_i=1]$ in $\theta_0$ can be consistently estimated by a moment estimator, $\sumi G_iY_{it}/\sumi G_i$. The second term in $\theta_0$ requires a regression model for the difference in crash counts $Y_{i,t+1}-Y_{it}$ given $\bfX_i$ among the control sites. Given that a regression model for the difference in counts is difficult to obtain, we separately assume a negative binomial model for each of the cross-sectional counts
\begin{eqnarray}\label{eq:nb}
\{Y_{it}(0)|\bfX_i,G_i=0\}&\sim &\NB(\mu(\bfX_i;\bfbeta),\phi), \nonumber\\
\{Y_{i,t+1}(0)|\bfX_i,G_i=0\}&\sim& \NB(\nu(\bfX_i;\bfgamma),\psi),
\end{eqnarray}
where $\mu$, $\nu$ are known smooth mean functions with parameter $\bfbeta$ and $\bfgamma$,  and the variances are $\mathbb{V}(Y_{it}(0)|\bfX_i,G_i=0)=\mu(\bfX_i;\bfbeta)+\mu^2(\bfX_i;\bfbeta)/\phi$ and $\mathbb{V}(Y_{i,t+1}(0)|\bfX_i,G_i=0)=\nu(\bfX_i;\bfgamma)+\nu^2(\bfX_i;\bfgamma)/\psi$, with potentially different dispersion parameters $\phi$ and $\psi$. Model (\ref{eq:nb}) is called the crash frequency model in traffic safety research \citep{AASHTO2010}. When the dispersion parameters approach infinity, model (\ref{eq:nb}) reduces to Poisson regression. As is evident from equation (\ref{eq:theta0}), the crash frequency model is only required for the control group, but not the treatment group. We obtain the maximum likelihood estimates of the parameters $\hat{\bfbeta}$ and $\hat{\bfgamma}$ using the control sites data. Under SUTVA and Assumption \ref{asp:pt}, equation (\ref{eq:theta0}) suggests the following estimator for $\theta_0$,
\begin{equation}\label{eq:reg}
\hat{\theta}_0^{\reg}=\frac{\sumi G_iY_{it}}{\sumi G_i}
+\frac{\sumi G_i\{\nu(\bfX_i;\hat{\bfgamma})-\mu(\bfX_i;\hat{\bfbeta})\}}{\sumi G_i}.
\end{equation}
When the crash frequency model (\ref{eq:nb}) is correctly specified, $\hat{\theta}^{\reg}_0$ is a consistent estimator of $\theta_0$, and thus $\hat{\tau}_{\CFD}^{\reg}=\hat{\theta}_1-\hat{\theta}_0^{\reg}$ and $\hat{\tau}_{\CMF}^{\reg}=\hat{\theta}_1/\hat{\theta}_0^{\reg}$ are consistent for $\tau_{\CFD}$ and $\tau_{\CMF}$, respectively.

The second estimator is based on weighting. Specifically, \citet{Abadie2005} showed that under Assumptions \ref{asp:pt} and weak overlap,
\begin{equation}\label{eq:mc}
\theta_0= \frac{1}{\pi}\E\Big\{G_iY_{it}+\frac{(1-G_i)(Y_{i,t+1}-Y_{it})e(\bfX_i))}{1-e(\bfX_i)}\Big\}.
\end{equation}
where $\pi=Pr(G_i=1)$.

If the propensity score is correctly estimated by $e(\bfX_i;\hat{\bfalpha})$, where $\bfalpha$ is the parameter of the propensity score model, equation \eqref{eq:mc} suggests the following weighting estimator for $\theta_0$:
\begin{equation}\label{eq:wt}
\hat{\theta}_0^{\wt}=\frac{\sumi G_iY_{it}w_i}{\sumi G_i}+
\frac{\sumi (1-G_i)(Y_{i,t+1}-Y_{it})w_i}{\sumi G_i},
\end{equation}
where $w_i=1$ for the treated group and $w_i=e(\bfX_i;\hat{\bfalpha})/[1-e(\bfX_i;\hat{\bfalpha})]$ for the control group. This further gives $\hat{\tau}_{\CFD}^\wt=\hat{\theta}_1-\hat{\theta}_0^{\wt}$ and $\hat{\tau}_{\CMF}^\wt=\hat{\theta}_1/\hat{\theta}_0^{\wt}$.
Re-weighting the observed crash counts by these ATT weights, we create a pseudo-population in which the covariates are balanced between treatment groups \citep*{Li2018}; the covariate balance consists the basis of valid group comparison. The weighting estimator avoids specifying the distributions of outcomes, but is in general not as efficient as outcome regression if the outcome model is correctly specified.

\subsection{Double-Robust Estimation}
The consistency of the regression estimator and the weighting estimator depends on the correct specification of the outcome model and propensity score model, respectively. Here, we propose a new hybrid DID estimator that augments weighting with regression:
\begin{equation}\label{eq:dr}
\hat{\theta}_0^{\dr}=\hat{\theta}_0^{\wt}
+\frac{1}{\sumi G_i}\sumi \frac{ (G_i-e(\bfX_i;\hat{\bfalpha}))\{\nu(\bfX_i;\hat{\bfgamma})-
\mu(\bfX_i;\hat{\bfbeta})\}}{1-e(\bfX_i;\hat{\bfalpha})}.
\end{equation}
This estimator can alternatively be written as a regression estimator augmented with weighting as
\begin{equation}\label{eq:dr2}
\hat{\theta}_0^{\dr}=\hat{\theta}_0^{\reg}
+\frac{\sumi (1-G_i)(\hat{R}_{i,t+1}-\hat{R}_{it})w_i}{\sumi G_i},
\end{equation}
where the residuals are defined as $\hat{R}_{i,t+1}=Y_{i,t+1}-\nu(\bfX_i;\hat{\bfgamma})$, $\hat{R}_{it}=Y_{it}-\mu(\bfX_i;\hat{\bfbeta})$. Based on these two equivalent formulations, we establish the following large-sample robustness property in Proposition \ref{prop:dr}, and include the proof in the Appendix.

\begin{proposition}\label{prop:dr}
As the sample size $n\rightarrow\infty$, the proposed estimator $\hat{\theta}_0^{\dr}$ converges in probability to $\theta_0$ if either $e(\bfX_i;\hat{\bfalpha})$ is consistent to the true propensity score or both $\nu(\bfX_i;\hat{\bfgamma})$ and $\mu(\bfX_i;\hat{\bfbeta})$ are consistent for the true mean functions.
\end{proposition}

By proposition \ref{prop:dr}, we can obtain the DR estimators for $\tau_{\CFD}$ and $\tau_{\CMF}$ by $\hat{\tau}_{\CFD}^{\dr}=\hat{\theta}_1-\hat{\theta}_0^{\dr}$ and $\hat{\tau}_{\CMF}^{\dr}=\hat{\theta}_1/\hat{\theta}_0^{\dr}$. In fact, estimator \eqref{eq:dr2} extends the DR estimator for ATT in the cross-sectional setting by \citet{Mercatanti2014}, who point out that the DR estimator may serve as a diagnostic tool in practical applications. Specifically, if the DR estimate differs from the regression estimate but is similar to the weighting estimate, it may suggest a potential misspecification of the regression function or lack of covariate overlap; if the DR estimate is close to the regression estimate but differs from the weighting estimate, it may suggest a potential misspecification of the propensity score model. We will exploit this diagnostic property of the DR estimate in the traffic safety study in Section \ref{sec:app}.

Although our presentation of the DR estimator is centered around the traffic safety application, the DR estimator should apply equally well in conventional program evaluation studies, such as estimating the causal effect of job training program on earnings \citep{Heckman1997,Heckman1998}. In that case, the causal estimand is usually defined on the additive scale similar to \eqref{eq:cfd}. However, since the earning outcomes are treated as continuous variables, the predicted mean functions $\nu$ and $\mu$ in the DR estimator could simply be obtained from the two-way fixed-effects model used in \citet{Ashenfelter1985} and \citet{Abadie2005} rather than from \eqref{eq:nb}.

For estimating the additive causal estimand, $\tau_{\CFD}$, in the traffic study, the DR estimator $\hat{\tau}_{\CFD}^{\dr}$ differs from the existing double-robust estimator for ATE, in the sense that $\hat{\theta}_0^{\dr}$ only requires estimating the outcome model among the control group but not the treated group. Further, the proposed estimator $\hat{\tau}^{\dr}_{\CFD}$ is indeed a member of the augmented inverse probability weighting (AIPW) estimator \citep{Robins1994}, but is distinct from the most efficient member, which necessarily requires an outcome model for the treated group. To obtain the most efficient AIPW-DID estimator, one could adapt the corresponding efficient AIPW estimator for estimating ATT designed for cross-sectional data \citep{Yang2018}, by essentially replacing their cross-sectional outcome with the before-after difference. Despite the efficiency advantage, we caution that such an estimator is not double-robust since it fails to be consistent to the target estimand once the propensity score model is misspecified. In traffic safety applications where the treated group often includes only a small number of pilot sites, the efficient DID estimator is less attractive because a count regression (e.g. negative binomial regression is routinely used in traffic safety studies) model tends to be unstable with non-convergence issues. For these reasons, we focus on the double-robust DID estimator $\hat{\theta}_0^{\dr}$ instead.

Since our estimator $\hat{\tau}_{\CFD}^{\dr}$ differs from the most efficient AIPW-DID estimator, we suspect that $\hat{\tau}_{\CFD}^{\dr}$ may not guarantee to be asymptotically more efficient than the weighting estimator when all models are correct. This is formalized in Proposition \ref{prop:eff} with an additive causal estimand and when the true propensity score is known. Proposition \ref{prop:eff} is not directly useful for inference as it assumes known propensity scores, but may provide insights for efficiency comparisons of the DID estimators in simulation experiments. Its proof is given in the Appendix.

\begin{proposition}\label{prop:eff}
For estimating the additive causal estimand, assuming the true propensity score is known, the $i$th influence function of the weighting estimator is
$$\varphi^{\text{wt}}_i=\frac{(G_i-e(\bfX_i))(Y_{i,t+1}-Y_{it})}{\pi(1-e(\bfX_i))}
-\tau_{\CFD}.$$
Further assuming the regression functions are known, the $i$th influence function of the double-robust estimator is
$$\varphi_i^{\dr}=\frac{(G_i-e(\bfX_i))(Y_{i,t+1}-Y_{it}-\{\nu(\bfX_i)-\mu(\bfX_i)\})}{\pi(1-e(\bfX_i))}
-\tau_{\CFD}.$$
The double-robust estimator is asymptotically at least as efficient as the weighting estimator only when $\mathbb{V}(\varphi_i^{\text{wt}})-\mathbb{V}(\varphi_i^{\text{dr}})\geq 0$. However, this inequality does not always hold. The full expression of $\mathbb{V}(\varphi_i^{\text{wt}})-\mathbb{V}(\varphi_i^{\text{dr}})$ is provided in the Appendix.
\end{proposition}

Even though the DR estimator is more robust to model misspecification in the DID design, Proposition \ref{prop:eff} suggests that it may be asymptotically less efficient than the weighting estimator even if all models are correctly specified. This is in sharp contrast to the existing results developed for estimating the average treatment effect (ATE). In the latter case, it is well-known that the double-robust estimator is asymptotically at least as efficient as the propensity score weighting estimator when all models are correctly specified \citep{Bang2005,Tsiatis2006}.

In the traffic safety application, we use a logistic regression to estimate the propensity scores. Since both the logistic and negative binomial models are smooth parametric models, we use the nonparametric bootstrap \citep{Efron1993} to obtain the associated $(1-\alpha)$ confidence interval (CI) and hence account for the uncertainty in estimating the nuisance parameters. For example, the following two steps are carried out to arrive at the CI estimator for $\hat{\tau}^{\dr}$. First, we re-sample with replacement from the empirical distribution $\hat{\mathbb{F}}_N(\mathcal{Z})$ to obtain the $b$th ($b=1,\ldots,B$) bootstrap replicate, $\{\mathcal{Z}_j^b,j=1,\ldots,N\}$, from which we compute $\hat{\tau}^{\dr,b}$. We then estimate the $\alpha/2$th and $(1-\alpha/2)$th quantiles of the collection of the bootstrap estimates, $\{\hat{\tau}^{\dr,b}, b=1,\ldots,B\}$, to form the lower and upper confidence limits for $\hat{\tau}^{\dr}$. Since $Y_{it}$ and $Y_{i,t+1}$ are repeated measurements from the same site in the before and after periods, there may be non-zero residual correlation between these crash counts. An advantage of the bootstrap procedure is that the correlation between repeated measurements are automatically taken into account by re-sampling the entire observed data vector $\mathcal{Z}_i$.

\section{Simulations}\label{sec:sim}
To illustrate the performance of different DID estimators, we conduct a small simulation study that mimics the real rumble strip application. Specifically, we simulate under a two-period two-group design. Each simulation has $N=2000$ units. Each unit has a binary covariate $X_1$ and a continuous covariate $X_2$, generated as follows:
\begin{equation*}
X_1 \sim \text{Bernoulli}(0.25),~~X_2|X_1 \sim \text{Normal}(2+6X_1,2^2).
\end{equation*}
We simulate the treatment group label $G_i$ independently from a Bernoulli distribution with success probability being the propensity score:
\begin{equation}\label{eq:pssim}
\text{logit}\{e(\bfX)\}=-2.0+X_1-0.2X_2+0.04X_2^2.
\end{equation}
Under the true propensity score model, the marginal treatment prevalence is approximately $20\%$, resembling our real application.

We generate the potential crash counts from negative binomial models, with different mean functions but same dispersion parameter $\phi=2.5$. Specifically, we assume \begin{equation*}
\begin{split}
Y_t(0)|\bfX,G=0 &\sim \text{NB}(\mu_{00}(\bfX),\phi),~~
Y_t(0)|\bfX,G=1 \sim \text{NB}(\mu_{01}(\bfX),\phi),\\
Y_{t+1}(0)|\bfX,G=0 &\sim \text{NB}(\nu_{00}(\bfX),\phi),~~
Y_{t+1}(1)|\bfX,G=1 \sim \text{NB}(\nu_{11}(\bfX),\phi),
\end{split}
\end{equation*}
with
\begin{equation}\label{eq:nbsim}
\begin{split}
\mu_{00}(\bfX)&=\exp(-2.0+0.4X_1+0.43X_2-0.022X_2^2),\\
\mu_{01}(\bfX)&=\exp(-3.0+0.3X_1+0.43X_2-0.022X_2^2),\\
\nu_{00}(\bfX)&=\exp(-1.9+0.5X_1+0.43X_2-0.022X_2^2),\\
\nu_{11}(\bfX)&=\exp(-2.5+0.1X_1+0.43X_2-0.022X_2^2).
\end{split}
\end{equation}
Under the parallel trend, the mean function of the counterfactual crash outcome for the treated sites is $\nu_{01}(\bfX)=\nu_{00}(\bfX)+\mu_{01}(\bfX)-\mu_{00}(\bfX)$. The coefficients of the mean functions in (\ref{eq:nbsim}) are informed by regression fit from analyzing the total crashes from the traffic safety application, and ensure that $\nu_{01}(\bfX)$ is positive over the support of $\bfX$. The true values of CFD and CMF, evaluated in large samples, are $-0.078$ and $0.862$, respectively.

We simulate $500$ replicates based on the models specified above. For each replicate, we use $\hat{\theta}_1$ to estimate $\theta_1$, but use different estimators for $\theta_0$. We first use the direct moment estimator based on the observed sample averages given in equation (\ref{eq:direct}). This estimator ignores pre-treatment covariate information and is only valid when there is no selection bias, namely, when the parallel trend holds unconditionally on the pre-treatment covariates. It is used here to quantify the selection bias in the data generation process. Further, the following estimators are compared.

\emph{Outcome regression}: we adopt the regression estimator in equation (\ref{eq:reg}) with correctly specified mean functions for $\mu(\bfX)$ and $\nu(\bfX)$. We also study the regression estimator with incorrectly specified mean functions that omit the linear term $X_1$ and the quadratic term $X_2^2$ in $\mu(\bfX)$ and $\nu(\bfX)$. These two estimators are labeled by REG and REG-mis, respectively.

\emph{Propensity score weighting}: we consider the weighting estimator in equation (\ref{eq:wt}) with the correctly specified propensity score model, as well as the weighting estimator with an incorrectly specified propensity score model that omits $X_1$ and $X_2^2$ in model \eqref{eq:pssim}. These two estimators are labeled by WT and WT-mis, respectively.

\emph{Double-Robust methods}: we compare the DR estimator in equation (\ref{eq:dr}) with correctly specified propensity score and outcome models (DR), the DR estimator with correctly specified outcome regression model but incorrectly specified propensity score model that omits $X_1$ and $X_2^2$ (DR-po), the DR estimator with correctly specified propensity score model but incorrectly specified outcome regression model that omits $X_1$ and $X_2^2$ (DR-ps), and the DR estimator with propensity score and outcome regression models being both incorrectly specified (DR-mis).

\begin{table}[htbp]
\centering
\caption{Absolute bias (Bias $\times 10^2$), root mean squared error (RMSE $\times 10^2$) and coverage of the $95\%$ confidence interval (Coverage) associated with each estimator for estimating $\tau_{\CFD}$ and $\log(\tau_{\CMF})$ in the simulations. The confidence intervals are computed based on 500 bootstrap samples from each simulated data set.} \label{tb:sim}\vspace{0.1in}
\scalebox{0.9}{
\begin{tabular}{lrrcrrc}
\toprule
& \multicolumn{3}{c}{$\tau_{\CFD}$} & \multicolumn{3}{c}{$\log(\tau_{\CMF})$} \\
\cmidrule(lr){2-4} \cmidrule(lr){5-7}
& Bias & RMSE & Coverage & Bias & RMSE & Coverage \\ \midrule
Direct & 13.4 & 14.5 & 33.4 & 27.6 & 30.5 & 38.4\\
REG & 0.4 & 13.4 & 94.8 & 1.9 & 26.6 & 94.8\\
REG-mis & 10.6 & 20.0 & 90.0 & 14.3 & 31.3 & 90.4 \\
WT & 0.2 & 14.1 & 95.6 & 2.6 & 27.7 & 95.6\\
WT-mis & 4.7 & 10.0 & 90.8 & 9.8 & 20.7 & 91.0 \\
DR & 0.5 & 14.5 & 95.4 & 2.2 & 28.6 & 95.4 \\
DR-po & 0.4 & 13.4 & 94.6 & 2.0 & 26.6 & 94.8 \\
DR-ps & 2.6 & 15.8 & 95.8 & 1.1 & 30.0 & 95.6\\
DR-mis & 7.0 & 16.7 & 91.8 & 9.2 & 27.6 & 92.0\\
\bottomrule
\end{tabular}}
\end{table}

Table \ref{tb:sim} presents the absolute bias, root mean squared error (RMSE) of each point estimator and the coverage of the corresponding $95\%$ bootstrap confidence intervals. Among all the estimators, the direct estimator is associated with the largest bias and RMSE and the lowest coverage in estimating both $\tau_{\CFD}$ and $\log(\tau_{\CMF})$. This is as expected because $X_1$ and $X_2$ affect both the treatment assignment and the potential outcomes, and induce selection bias. The DID regression, weighting, and DR estimators all present small and comparable bias when the corresponding models are correctly specified. When the outcome regression functions $\mu(\bfX)$ and $\nu(\bfX)$ are misspecified, the regression estimator shows inflated bias and RMSE, with reduced coverage. Similarly, misspecification of the propensity score model also leads to increased bias and sub-nominal coverage for the DID weighting estimator. In this simulation, the substantial reduction in the variance of the weights from a misspecified propensity score model appears to outweigh the inflation in bias, which explains the decreased RMSE associated with WT-mis relative to WT.

The simulation also demonstrates the double robustness property of the DID-DR estimator: when either the propensity score model or outcome model is misspecified, the DR estimator (DR-po and DR-ps) leads to small bias and nominal coverage for both estimands. Interestingly, in estimating the additive effect $\tau_{\CFD}$, the outcome model appears have a bigger impact on the DR estimator than the propensity score model. Specifically, when only the outcome model is correctly specified, the DR estimator performs very close to the DR estimator with both models being correctly specified, but the DR estimator under-performs much if only the propensity score is correct. Similar phenomenon was previously observed in the DR estimation of ATT and ATE in the cross-sectional setting \citep*[e.g.][]{Li2013}. This pattern is not obvious for ratio estimand $\log(\tau_{\CMF})$, likely because that the bias---an additive and scaled quantity by definition---of a scale-free ratio quantity does not fully capture the true discrepancy in estimating $\theta_1$ and $\theta_0$. Lastly, when both the propensity score and outcome models are misspecified, the DR estimator (DR-mis) results in inflated bias and under-coverage; nonetheless, even under this scenario, the misspecified DR estimator still outperforms the corresponding misspecified regression estimator with 46\% and 6\% reduction in relative bias for estimating the additive and ratio estimands, respectively. In addition, we observe in the simulations that the Monte Carlo variance of the DR estimator, when both models are correctly specified, is very close to that of the weighting estimator with a correct propensity score model. This phenomenon may be partially explained by Proposition \ref{prop:eff}. Specifically, under the current data generating process mimicking the traffic safety application, we found that the Monte Carlo estimate of $N^{-1}\{{\mathbb{V}}(\varphi_i^{\text{wt}})-{\mathbb{V}}(\varphi_i^{\text{dr}})\}<0$ is negative and close to zero (averaged across simulation iterations).

\section{Application to the Pennsylvania Rumble Strip Data}\label{sec:app}
\subsection{The Data}
Our application is based on the Federal Highway Administration Evaluation of Low-Cost Safety Improvements Pooled Fund Study \citep*{Lyon2015}. The study embraced a broader scope and focused on quantifying the safety effectiveness of the combined application of centerline and shoulder rumble strips in mitigating crash outcomes among two-lane rural road locations in Kentucky, Missouri, and Pennsylvania. We obtained the subset of traffic safety records from the Pennsylvania Department of Transportation (PennDOT; \url{http://www.penndot.gov/}), which includes vehicle crash counts for traffic sites within the state of Pennsylvania up to 2012. Since each traffic site is a roadway segment, we use these two terms interchangeably. From $2009$ to $2011$, centerline and shoulder rumble strips were installed in $331$ rural, undivided two-lane roadway segments for a total of over $200$ miles. The control group consists of five times more sites that did not receive rumble strips before $2012$ but had similar traffic volume. Therefore, the data we analyze consist of around $2000$ rural highway segments, approximately $17\%$ of which received the treatment. We define year $2008$ as the before period and year $2012$ as the after period.

We consider four types of crash outcomes: (1) fatal-plus-injury (FI)---crashes that involve at least one fatal or injured person; (2) property-damage-only (PDO)---crashes where no occupant was injured; (3) run-off-the-road (ROR)---crashes where a vehicle travels outside the trafficway and collides with a natural or artificial object in an area not intended for vehicles; this is a subset of the first two crash types; (4) total number of fatal-plus-injury and property-damage-only crashes (TOT). Table \ref{tb:crash} presents the aggregated crash counts for each type among both treated and control sites in the before and after periods.

\begin{table}[htbp]
\centering
\caption{Crash counts by type for both treated and control sites in the before and after periods. FI: fatal-plus-injury; PDO: property-damage-only; ROR: run-off-the-road; TOT: total.}\label{tb:crash}\vspace{0.1in}
\scalebox{0.9}{
\begin{tabular}{lcccc}
\toprule
& \multicolumn{2}{c}{Treated ($N_1=331$)} & \multicolumn{2}{c}{Control ($N_0=1,655$)} \\
\cmidrule(lr){2-3} \cmidrule(lr){4-5}
 & Before & After & Before & After \\ \midrule
FI & 78 & 77 & 441 & 436 \\
PDO & 61 & 41 & 350 & 321\\
ROR & 22 & 21 & 123 & 143 \\
TOT & 139 & 118 & 791 & 757\\
\bottomrule
\end{tabular}}
\end{table}

The pre-treatment covariates we consider are site-specific characteristics often suggested in constructing crash frequency models \citep{AASHTO2010}. These variables include the operational characteristic of a roadway segment, the speed limit (high speed if the posted limit is above 45 mph and low speed otherwise), as well as geometric features of a roadway: segment length in miles, pavement width (three categories), average shoulder width (three categories), number of driveways (three categories), existence of intersections (two categories), existence of curves (two categories) and average degree of curvature. An important covariate is AADT---the average annual daily traffic volume. Although strictly speaking AADT is a time-varying covariate, we found that in this application the AADT of the before and after periods are very similar across all sites; thus we assume AADT is time-invariant and take the before period value as the covariate. Table \ref{tb:des} presents the descriptive statistics of the covariates.

\begin{table}[htbp]
\centering
\caption{Definition of variables and their descriptive statistics by treatment group. Mean (st. dev), [Min, Max] values are given for each continuous variable and the number of traffic sites (percentages) are given for each level of the categorical variables. Total sample size $N=1,986$.}\label{tb:des}\vspace{0.1in}
\scalebox{0.8}{
\begin{tabular}{llcc}
\toprule
Variable & Definition & Treated & Control \\ \midrule
AADT & Annual average daily traffic & 3,520 (2,628) & 3,636 (2,495) \\\smallskip
& volume; vehicles per day & [818, 15,033] & [678, 15,379]\\
Length & Roadway segment length & 0.47 (0.16) & 0.48 (0.13)\\\smallskip
& in miles & [0.01, 0.75] & [0.03, 0.76]\\
Width & Pavement width in feet & & \\
& \quad width $\leq 20$ & 20 (6.0) & 346 (20.9)\\
& \quad $20<$ width $\leq 23$ & 169 (51.1) & 828 (50.0)\\\smallskip
& \quad otherwise & 142 (42.9) & 481 (29.1)\\
Speed & Posted speed limit & & \\
& \quad low if limit $\leq 45$ mph & 216 (65.3) & 956 (57.9)\\\smallskip
& \quad high otherwise & 115 (34.7) & 696 (42.1)\\
Shoulder & Average shoulder width in feet & &\\
& \quad width $\leq 3$ & 88 (26.6) & 867 (52.4)\\
& \quad $3<$ width $\leq 6$ & 191 (57.7) & 643 (38.8)\\\smallskip
& \quad otherwise & 52 (15.7) & 145 (8.8)\\
Driveways & Number of driveways & & \\
& \quad no driveway & 24 (7.2) & 101 (6.1)\\
& \quad $1\leq $ number of driveways $\leq 10$ & 235 (71.0) & 1,100 (66.5)\\\smallskip
& \quad otherwise & 72 (21.8) & 454 (27.4)\\
Intersections & Inclusion of intersections & &\\
& \quad No intersections & 257 (77.6) & 1,162 (70.2)\\\smallskip
& \quad At least $1$ intersection & 74 (22.4) & 493 (29.8)\\
Curves & Existence of horizontal curves & & \\
& \quad No curves & 143 (43.2) & 701 (42.4)\\\smallskip
& \quad At least $1$ curve & 188 (56.8) & 954 (57.6)\\
Curvature & Average degree of curvature & 2.81 (3.59) & 3.92 (7.59)\\
& & [0, 25.28] & [0, 132.30] \\
\bottomrule
\end{tabular}}
\end{table}

\subsection{Model Specification}
We estimate the propensity score by logistic regression including all the pre-treatment site characteristics. We adopt the power series specification for the continuous variables (AADT, Length and Curvature) with the optimal order of terms $1\leq l\leq 5$ selected by leave-one-out-cross-validation. We choose to include up to the third-order terms of the continuous variables in the propensity score model since $l=3$ corresponds to the lowest mean squared error for predicting treatment. The fitted propensity score model suggests that road segments with wider pavement and shoulder, low speed limit, at least one driveway, no intersections nor curves are more likely to receive rumble strip installation.

For both the regression and DR estimators, we model the cross-sectional means of potential outcomes among the reference sites during each period. AADT and segment length are transformed to the log scale, as is common practice in developing crash frequency models in traffic safety research \citep{AASHTO2010}. To allow for over-dispersion, we use negative binomial regression to estimate model parameters. Specifically, we assume the conditional distributions of $Y_{it}(0)$ and $Y_{i,t+1}(0)$ given $\bfX_i$ in the reference group as in (\ref{eq:nb}), where
\begin{equation}\label{eq:spf}
\begin{split}
\mu(\bfX)&=L^{\beta_L}\cdot \text{AADT}^{\beta_{\text{AADT}}}\cdot\exp\Big(\beta_0+\sum_{j=1}^J\beta_jX_j\Big),\\
\nu(\bfX)&=L^{\gamma_L}\cdot \text{AADT}^{\gamma_{\text{AADT}}}\cdot\exp\Big(\gamma_0+\sum_{j=1}^J\gamma_jX_j\Big).
\end{split}
\end{equation}
In (\ref{eq:spf}), $L$ denotes the segment length, AADT is the traffic volume and $J$ is the number of remaining covariates (including dummy variables). We adopt the log-linear specification for the outcome model since it performs as well as its power series counterpart regarding mean squared error estimated by leave-one-out-cross-validation, and yet is computationally convenient without convergence issues.

\subsection{Assessment of Overlap, Balance and Parallel Trend}\label{sec:assess}
\begin{figure}[htbp]
\includegraphics[scale=0.19]{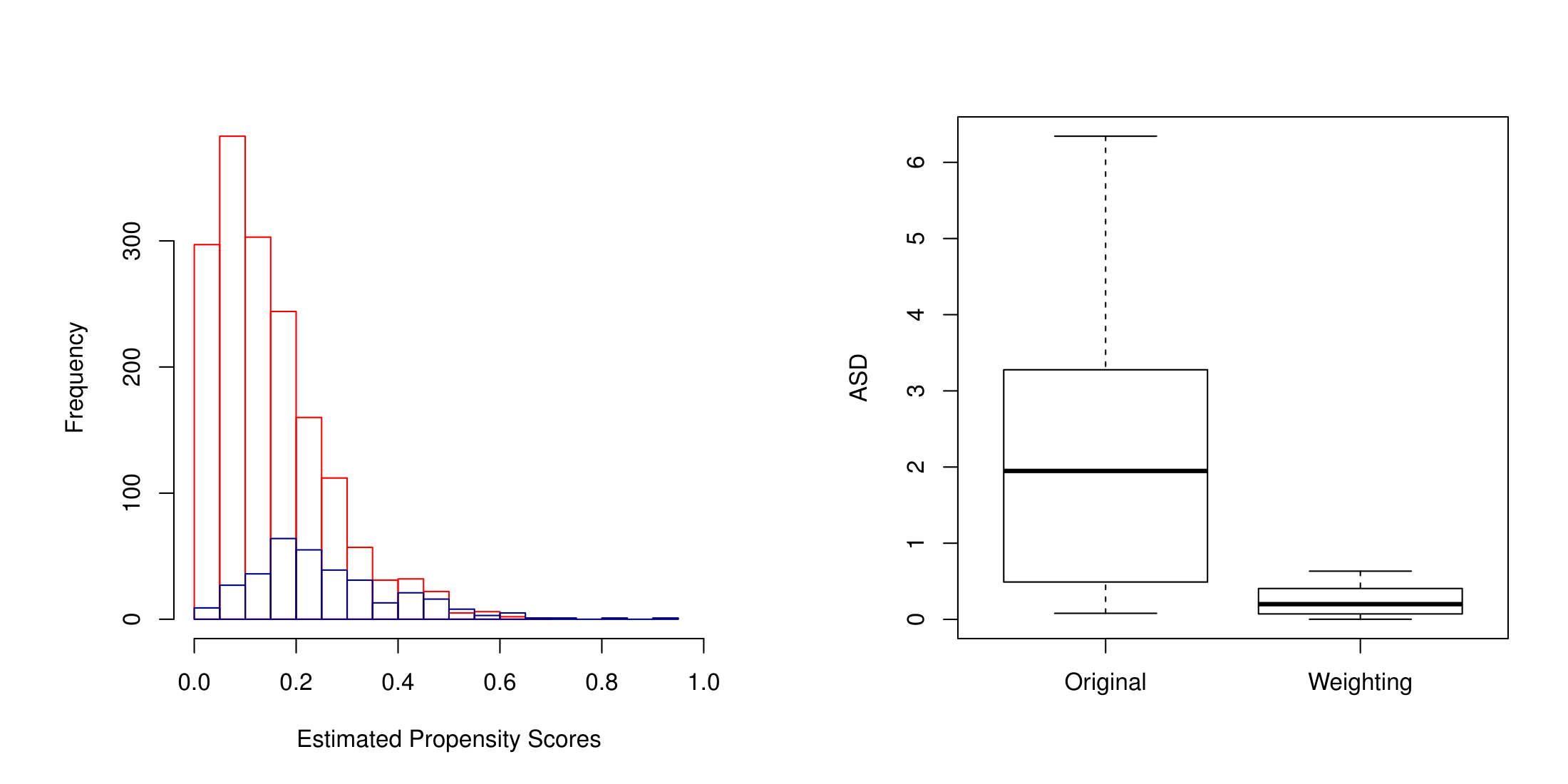}
\caption{Left panel: histogram of the estimated propensity score for the treated sites (blue) and the control sites (red). Right panel: boxplot of the absolute standardized difference all covariates in the original and weighted data.}
\label{fig:Fig1}
\end{figure}

We  assess the weak overlap assumption by visually checking the overlap in the histograms of the estimated propensity scores for the treated and control sites (Figure \ref{fig:Fig1}). The histogram suggests satisfactory overlap between the two groups.

We further check the covariate balance in the original and weighted sample by calculating the absolute standardized difference (ASD) of each covariate (including up to the third-order term for each continuous variable) between the two treatment groups, defined as
\begin{equation}
\text{ASD}=\left\vert\frac{\sumi G_iX_iw_i}{\sumi G_iw_i}-\frac{\sumi (1-G_i)X_iw_i}{\sumi (1-G_i)w_i}\right \vert \Big/ \sqrt{\frac{s_1^2}{N_1}+\frac{s_0^2}{N_0}},
\end{equation}
where $N_1$, $N_0$ are the number of treated and reference sites, $s_1^2$, $s_0^2$ are the variances of the unweighted covariate in the treated and control group, respectively. The weight $w_i=1$ for all sites in the original data and the ASD is the standard two-sample $t$-statistic. For the weighted data, $w_i$ is the ATT weight introduced in Section \ref{sec:exist}. The right panel in Figure \ref{fig:Fig1} presents the boxplots of the ASD; it shows that propensity score weighting substantially improves the covariate balance, with the largest ASD value equal to $0.63$ in the weighted sample compared to $6.34$ in the unweighted sample (the standard threshold for significant difference is $1.96$). The good covariate balance supports that the propensity scores are well estimated.

\begin{table}[h!]
\centering
\caption{Estimated CFD ($\hat{\tau}_{\CFD}$) and CMF ($\hat{\tau}_{\CMF}$) and the corresponding $95\%$ confidence intervals for all crash types in the ``no treatment" evaluation using different DID estimators.}\label{tb:lagged}\vspace{0.1in}
\scalebox{0.8}{
\begin{tabular}{llcccc}
\toprule
 &  & Direct & REG & WT & DR \\ \midrule
{FI} & {$\hat{\tau}_{\CFD}$} & -0.060 & -0.026 & -0.029 & -0.028\\
&  & (-0.144,0.026) & (-0.104,0.062) & (-0.121,0.073) & (-0.123,0.067)\\
& {$\hat{\tau}_{\CMF}$} & 0.798 & 0.902 & 0.891 & 0.892 \\\smallskip
&  & (0.587,1.107) & (0.651,1.331) & (0.621,1.370) & (0.623,1.341)\\
{PDO} & {$\hat{\tau}_{\CFD}$} & 0.008 & 0.010 & 0.026 & 0.027\\
&  & (-0.066,0.078) & (-0.066,0.086) & (-0.052,0.108) & (-0.052,0.110)\\
& {$\hat{\tau}_{\CMF}$} & 1.045 & 1.059 & 1.164 & 1.169\\\smallskip
&  & (0.692,1.594) & (0.693,1.686) & (0.738,2.054) & (0.744,2.106)\\
{ROR} & {$\hat{\tau}_{\CFD}$} & -0.016 & 0.001 & 0.007 & 0.009\\
&  & (-0.064,0.029) & (-0.051,0.046) & (-0.043,0.053) & (-0.044,0.054)\\
& {$\hat{\tau}_{\CMF}$} & 0.809 & 1.014 & 1.121 & 1.150 \\\smallskip
&  & (0.431,1.522) & (0.491,2.565) & (0.524,3.096) & (0.521,3.234) \\
{TOT} & {$\hat{\tau}_{\CFD}$} & -0.052 & -0.015 & -0.003 & -0.002 \\
&  & (-0.146,0.073) & (-0.117,0.109) & (-0.119,0.140) & (-0.115,0.138)\\
& {$\hat{\tau}_{\CMF}$} & 0.890 & 0.965 & 0.993 & 0.996 \\
&  & (0.714,1.192) & (0.761,1.333) & (0.764,1.427) & (0.762,1.425)\\
\bottomrule
\end{tabular}}
\end{table}

To indirectly assess the key parallel trend assumption, we perform a DID analysis of the crash outcomes for two pre-treatment periods. Specifically, we obtain the crash outcome, $Y_{i,t-1}$, during the year of 2004 for each traffic site and treat it as the proxy-before observation; the crash outcome, $Y_{it}$, during the year of 2008 are then regarded as the proxy-after data. As discussed in Section \ref{sec:aspt}, if the parallel trend assumption is plausible, then the estimated CFD and CMF based on the proxy-before-after observations should be close to $0$ and $1$, respectively, because in reality rumble strips were not applied until after $2008$.

Table \ref{tb:lagged} presents the results of this ``no treatment" analysis. For all crash types, DR and weighting estimators produce similar estimates for CFD and CMF. Overall, the confidence intervals for CFD include $0$ for all types of crashes regardless of the choice of DID estimator. However, it is worth noting that the CFD estimates from DR and weighting for the ROR and total crashes are close to $0$, which further support the plausibility of the parallel trend. By contrast, there is a potential for violation of parallel trend regarding FI and PDO crashes since the DR estimates for CFD tend to deviate from the null. Nevertheless, the lack of statistical significance may still permit the subsequent DID analyses.

\begin{table}[h!]
\centering
\caption{Estimated CFD ($\hat{\tau}_{\CFD}$) and CMF ($\hat{\tau}_{\CMF}$) and the corresponding $95\%$ confidence intervals for all crash types with before and after data using different DID methods.}\label{tb:res}\vspace{0.1in}
\scalebox{0.8}{
\begin{tabular}{llcccc}
\toprule
 & & Direct & REG & WT & DR \\ \midrule
{FI} & {$\hat{\tau}_{\CFD}$} & 0.000 & -0.022 & -0.009 & -0.008\\
&  & (-0.087,0.078) & (-0.118,0.060) & (-0.110,0.077) & (-0.108,0.079)\\
& {$\hat{\tau}_{\CMF}$} & 1.000 & 0.912 & 0.963 & 0.966 \\\smallskip
&  & (0.706,1.470) & (0.629,1.318) & (0.657,1.458) & (0.660,1.474)\\
{PDO} & {$\hat{\tau}_{\CFD}$} & -0.043 & -0.037 & -0.056 & -0.058\\
&  & (-0.113,0.026) & (-0.106,0.036) & (-0.134,0.022) & (-0.137,0.018)\\
& {$\hat{\tau}_{\CMF}$} & 0.743 & 0.770 & 0.687 & 0.681 \\\smallskip
&  & (0.455,1.223) & (0.462,1.384) & (0.409,1.196) & (0.404,1.180)\\
{ROR} & {$\hat{\tau}_{\CFD}$} & -0.015 & -0.022 & -0.039 & -0.039 \\
&  & (-0.060,0.029) & (-0.066,0.022) & (-0.099,0.006) & (-0.096,0.006) \\
& {$\hat{\tau}_{\CMF}$} & 0.808 & 0.746 & 0.617 & 0.619 \\\smallskip
& & (0.437,1.592) & (0.417,1.382) & (0.328,1.085) & (0.331,1.090) \\
{TOT} & {$\hat{\tau}_{\CFD}$} & -0.043 & -0.060 & -0.065 & -0.066 \\
&  & (-0.145,0.063) & (-0.174,0.049) & (-0.188,0.052) & (-0.191,0.052) \\
& {$\hat{\tau}_{\CMF}$} & 0.893 & 0.856 & 0.845 & 0.844 \\
&  & (0.684,1.189) & (0.651,1.154) & (0.627,1.166) & (0.626,1.154)\\
\bottomrule
\end{tabular}}
\end{table}

\subsection{Results}
We analyzed crash outcomes in 2008 and 2012 using different DID estimators for all crash types and present the results in Table \ref{tb:res}. As observed in the simulations, the direct estimator (\ref{eq:direct}) is subject to selection bias and tends to give different results from the rest. Across all four crash types, the DR estimator produces CFD and CMF results similar to the weighting estimator, but sometimes different from the regression estimator. Given the satisfactory overlap indicated in Figure \ref{fig:Fig1}, the difference in estimates suggests that the outcome regression model may be mildly misspecified. The CFD and CMF for FI crashes estimated by the DR approach are both close to the null values, implying negligible effect from rumble strips on mitigating FI crashes. The application of rumble strips seems to reduce the PDO crashes with CFD and CMF estimated to be $\hat{\tau}_{\CFD}^{\dr}=-0.058$ and $\hat{\tau}_{\CMF}^{\dr}=0.681$ using the DR approach. However, cautions need to be exercised to interpret these estimates since the parallel trend assumption may be questionable, as discussed previously. The parallel trend is deemed plausible for the total crashes, and we find that rumble strips have a potentially beneficial effect on total crash frequency ($\hat{\tau}_{\CFD}^{\dr}=-0.066$ and $\hat{\tau}_{\CMF}^{\dr}=0.844$), but the $95\%$ CIs include the null values. Additionally, the application of rumble strips suggests a potential causal effect on mitigating the ROR crashes, with $\hat{\tau}_{\CFD}^{\dr}=-0.039$ and $\hat{\tau}_{\CMF}^{\dr}=0.619$ estimated by the DR approach, but the CIs cover $0$ and $1$. Overall, our analysis only finds beneficial but statistically insignificant effects of rumble strips on reducing crashes. This agrees with the empirical findings of several other traffic safety studies based on alternative data sources and modeling strategies \citep{Griffith1999,Garder2006,Khan2015}.

\section{Discussion}\label{sec:dis}
In this paper, we draw causal inference in traffic safety before-after studies within the DID framework and propose a new double-robust DID estimator. The primary concern for observational traffic safety data is related to bias, which may be due to confounding, site selection or model misspecification, among others. Our DR estimator grants two chances for consistent estimation of the causal effect and has been demonstrated to have small bias from misspecification of either the propensity score model or the outcome model. Applying the DR method and several alternative methods to a real data, we find that rumble strips have a moderate but statistically insignificant beneficial effect in reducing vehicle crashes. These insignificant findings may be partially due to the limited number of crash events over a one-year period, a limitation of our available data. It would be of interest to update the CFD and CMF estimates with longer before and after periods.

Though the causal rate ratio estimand, CMF, dominates the traffic safety studies, we recommend assessing alternative estimand such as the causal rate difference, CFD to offer a more comprehensive picture of the effectiveness. This is because that CMF is scale-free and does not inform the absolute change in the expected crash frequency. For example, in our application, the CFD estimate suggests a modest absolute change in crash frequency ($\hat{\tau}_{\CFD}^{\dr}=-0.039$) for the ROR crashes, which can be translated into an average reduction of $4$ crashes per $100$ road segments due to rumble strips. On the other hand, the CMF estimate is $\hat{\tau}_{\CMF}^{\dr}=0.617$, which indicate a large proportional change. This slight discrepancy comes from the fact that the ROR crashes constitute a small fraction of the total crashes.

There are several limitations of this study. First, a limitation of the DID framework is that the parallel trend assumption is scale-dependent. For example, it may hold for the original $Y$ but not for a nonlinear monotone transformation of $Y$, such as $\log(Y)$. A common alternative scale-free identification condition for the before-after design is the ignorability assumption conditional on the lagged outcomes. In the context of linear models, \citet{Angrist2009} show that the DID estimate and lagged-outcome regression estimates have a bracketing relationship. Namely, if ignorability is correct, then mistakenly assuming parallel trend will overestimate a true positive effect; in contrast, if parallel trend is correct, then mistakenly assuming ignorability will underestimate a true positive effect. Thus, one can treat the estimate under each assumption as the upper and lower bounds of the true effect in practice. It is particularly relevant to traffic safety studies---where the outcome is usually counts---to evaluate whether such a bracketing relationship holds more generally beyond the linear setting.

Second, the SUTVA may be violated and such a violation could lead to a biased average causal effect estimate. The violation is more likely if the traffic sites are adjacent to each other allowing for a potential spillover effect. For example, it is possible that a drowsy and fatigued driver was alerted in a roadway segment with shoulder rumble strips and hence was less likely to have a run-off-the-road crash in a nearby reference site, thus biasing the causal estimate towards the null. It is also likely that \emph{crash migration} leads to violation of SUTVA. For instance, a vehicle travelling through a reference site with low visibility may end up in a crash in a consecutive site with rumble strips. However, the reporting officer usually traced the location where the crash was initiated by a careful analysis of the available evidence at the crash site, and may mitigate such concerns. In any case, the extension of the DID analysis that accounts for interference between roadway segments in the spirit of \citet{Hudgens2008} would be of interest.

We have adopted smooth parametric models to estimate the propensity score and the crash counts. In this case, the resulting DID estimators are all asymptotically linear, and the nonparametric bootstrap enables valid inference \citep{Shao2012}. This also underlies why the bootstrap CI maintains nominal coverage for the DR-po and DR-ps estimators in our simulation study. In practice, since well-estimated propensity score and outcome models are critical for the consistency of the DR estimator, an appealing strategy is to leverage data-adaptive machine learning techniques for estimating the propensity scores and for predicting the crash counts. Specifically, one could use boosting to estimate the propensity score \citep{McCaffrey2004,McCaffrey2013}, which has been demonstrated to maintain adequate weighted covariate balance \citep*{Lee2009}, or use random forest to better predict the counterfactual safety outcomes \citep{Breiman2001,Liaw2002}. However, the nonparametric bootstrap CI may not guarantee to carry nominal coverage in those cases since the resulting estimator may no longer be asymptotically linear.

Finally, we have only developed double-robust estimation within the canonical two-period DID design with panel data. More complicated before-after data structure may arise in other policy evaluation contexts. For example, the treatment (policy) could be administered to a small number of states, and repeated cross sections or surveys are taken at the household level or individual level to measure the before and after outcomes. When repeated cross sections or random surveys are taken in both the before and after periods (rather than panel observations for the same group of units), the proposed double-robust DID estimator may not directly apply since the identification condition differs from equation \eqref{eq:mc}. \citet{Abadie2005} provided the revised identification condition and suggested the corresponding weighting estimator (Section 3.2.1 of the paper). Therefore, an appropriate double-robust estimator is obtained by modifying the propensity score weighting component along the lines of \citet{Abadie2005}. Additionally, one must address the within-state correlations among households or individuals when the treatment is applied at the state level. In particular, valid bootstrap should proceed by resampling the states so that the within-state correlation structure is preserved \citep{Li2013}. In other program evaluation applications with staggered adoption and multiple time periods, the two-way fixed-effects model represents a standard regression estimator for causal inference. \citet{Callaway2018} recently defined several new average causal estimands appropriate for the multiple-period DID design, and studied a propensity score weighting estimator. An important avenue for future research is to provide a double-robust DID extension by combining the weighting estimator of \citet{Callaway2018} and the two-way fixed-effects outcome model for improved inference with multiple-period data.

\section*{Appendix}
\subsection*{Proof of Proposition \ref{prop:dr}}
The DR estimators are constructed as $\hat{\tau}_{\CFD}^{\dr}=\hat{\theta}_1-\hat{\theta}_0^{\dr}$ and $\hat{\tau}_{\CMF}^{\dr}=\hat{\theta}_1/\hat{\theta}_0^{\dr}$; the moment-based estimator
\begin{equation}\label{eq:a1}
\hat{\theta}_1=\frac{\sumi G_iY_{i,t+1}}{\sumi G_i}\cp\frac{\E[G_iY_{i,t+1}]}{\pi}=\E[Y_{i,t+1}(1)|G_i=1]=\theta_1,
\end{equation}
where $\pi=Pr(G_i=1)>0$. To show that $\hat{\tau}_{\CFD}^{\dr}$ and $\hat{\tau}_{\CMF}^{\dr}$ are double-robust for estimating $\tau_{\CFD}$ and $\tau_{\CMF}$, it suffices to show that $\hat{\theta}_0^{\dr}$ is double-robust for estimating $\theta_0$.

We first assume that the propensity score model $e(\bfX;\bfalpha)$ is correctly specified while the outcome model may be subject to misspecification. We assume certain regularity conditions hold (e.g., smooth regression functions and bounded moments for all covariates), and denote the maximum likelihood estimators for model parameters by $\hat{\bfalpha}$, $\hat{\bfgamma}$ and $\hat{\bfbeta}$. Under these assumptions, $\hat{\bfalpha}\cp \bfalpha_0$, $\hat{\bfgamma}\cp \bfgamma^*$, $\hat{\bfbeta}\cp \bfbeta^*$, where $\bfalpha_0$ is the true value for the correct propensity score model but $\bfgamma^*$, $\bfbeta^*$ may be different from the true values $\bfgamma_0$, $\bfbeta_0$. By the results of \citet{White1982}, $\bfgamma^*$ and $\bfbeta^*$ are the least false values that minimize the Kullback-Leibler distance between the probability distribution based on the postulated models and the true data generating models. We first observe that the last term on the right hand side of equation (\ref{eq:dr}) converges in probability to zero. To see why, we write
\begin{equation*}
\begin{split}
&\frac{1}{\sumi G_i}\sumi \frac{ (G_i-e(\bfX;\hat{\bfalpha}))\{\nu(\bfX_i;\hat{\bfgamma})-\mu(\bfX_i;\hat{\bfbeta})\}}
{1-e(\bfX;\hat{\bfalpha})}\\
\cp& \frac{1}{\pi}\E\Big[\frac{ (G_i-e(\bfX;\bfalpha_0))\{\nu(\bfX_i;\bfgamma^*)-\mu(\bfX_i;\bfbeta^*)\}}
{1-e(\bfX;\bfalpha_0)}\Big]\\
=&\frac{1}{\pi}\E\Big[\frac{ \{\E(G_i|\bfX_i)-e(\bfX;\bfalpha_0)\}\{\nu(\bfX_i;\bfgamma^*)-\mu(\bfX_i;\bfbeta^*)\}}
{1-e(\bfX;\bfalpha_0)}\Big]=0,
\end{split}
\end{equation*}
where the second to last equation is an application of the Law of Iterated Expectation. Therefore by (\ref{eq:dr}), it is immediate that $\hat{\theta}_0^{\dr}$ shares the same probability limit with $\hat{\theta}_0^{\wt}$, which is consistent to $\theta_0$ when the propensity score model is correctly specified \citep{Abadie2005}. This is why $\hat{\theta}_0^{\dr}\cp \theta_0$.

Alternatively, suppose the outcome model is correctly specified but the propensity score model may subject to misspecification. In this case, $\hat{\bfgamma}\cp \bfgamma_0$, $\hat{\bfbeta}\cp \bfbeta_0$, $\hat{\bfalpha}\cp \bfalpha^*$, where $\bfalpha^*$ minimizes the Kullback-Leibler distance between the probability distribution based on the postulated model and the true data generating model \citep{White1982} and thus may differ from truth data generating model parameter $\bfalpha_0$. Then the last term on the right hand side of (\ref{eq:dr2}) converges in probability to zero. This is because
\begin{equation*}
\begin{split}
&\frac{\sumi (1-G_i)(\hat{R}_{i,t+1}-\hat{R}_{it})w_i}{\sumi G_i}\\
=&\frac{\sumi (1-G_i)}{\sumi G_i} \frac{1}{\sumi (1-G_i)}\sumi \frac{(1-G_i)
(\hat{R}_{i,t+1}-\hat{R}_{it})e(\bfX;\hat{\bfalpha})}{1-e(\bfX;\hat{\bfalpha})}\\
\cp &\frac{1-\pi}{\pi}\E\Big[\frac{
(Y_{i,t+1}-Y_{it})e(\bfX;\bfalpha_0)}{1-e(\bfX;\bfalpha_0)}\Big|G_i=0\Big]
-\frac{1-\pi}{\pi}\E\Big[\frac{
\{\nu(\bfX_i;\bfgamma_0)-\mu(\bfX_i;\bfbeta_0)\}e(\bfX_i)}{1-e(\bfX;\bfalpha_0)}\Big|G_i=0\Big].\\
\end{split}
\end{equation*}
and
\begin{equation*}
\begin{split}
&\Big[\frac{
(Y_{i,t+1}-Y_{it})e(\bfX;\bfalpha_0)}{1-e(\bfX;\bfalpha_0)}\Big|G_i=0\Big]\\
=&\E\Big[\frac{
\{Y_{i,t+1}(0)-Y_{it}(0)\}e(\bfX;\bfalpha_0)}{1-e(\bfX;\bfalpha_0)}\Big|G_i=0\Big]\\
=&\E\Big[\frac{e(\bfX;\bfalpha_0)}{1-e(\bfX;\bfalpha_0)}\E[Y_{i,t+1}(0)-Y_{it}(0)|\bfX_i,G_i=0]\Big|G_i=0\Big]\\
=&\E\Big[\frac{e(\bfX;\bfalpha_0)}{1-e(\bfX;\bfalpha_0)}\{\nu(\bfX_i;\bfgamma_0)-
\mu(\bfX_i;\bfbeta_0)\}\Big|G_i=0\Big],
\end{split}
\end{equation*}
where the second to last equality is granted by the Law of Iterated Expectation and the last equality comes from the definition of the regression function. By (\ref{eq:dr2}), it follows that $\hat{\theta}_0^{\dr}$ shares the same probability limit with $\hat{\theta}_0^{\reg}$, which is consistent to $\theta_0$ when the cross-sectional crash frequency model is correctly specified. Therefore $\hat{\theta}_0^{\dr}\cp \theta_0$, and the double-robust property holds.

\subsection*{Proof of Proposition \ref{prop:eff}}
Denote the $i$th known true propensity score as $e(\bfX_i)$, then the weighting estimator is
\begin{equation*}
\begin{split}
\tau_{\CFD}^{\wt}&=\frac{\sumi G_i(Y_{i,t+1}-Y_{it})}{\sumi G_i}-
\frac{1}{\sumi G_i}\sumi \frac{(1-G_i)(Y_{i,t+1}-Y_{it})e(\bfX_i)}{1-e(\bfX_i)}\\
&=\left(\frac{N}{\sumi G_i}\right)
\left\{\frac{1}{N}\sumi G_i(Y_{i,t+1}-Y_{it})-
\frac{1}{N}\sumi \frac{(1-G_i)(Y_{i,t+1}-Y_{it})e(\bfX_i)}{1-e(\bfX_i)}\right\}.
\end{split}
\end{equation*}
Further observe that
\begin{equation*}
\begin{split}
&\sqrt{N}(\tau_{\CFD}^{\wt}-\tau_{\CFD})\\
=&\frac{1}{\pi}\frac{1}{\sqrt{N}}\sumi
\left\{G_i(Y_{i,t+1}-Y_{it})-
\frac{(1-G_i)(Y_{i,t+1}-Y_{it})e(\bfX_i)}{1-e(\bfX_i)}-\tau_{\CFD}\right\}+o_p(1)\\
=&\frac{1}{\pi}\frac{1}{\sqrt{N}}\sumi
\left\{\frac{(G_i-e(\bfX_i))(Y_{i,t+1}-Y_{it})}{\pi(1-e(\bfX_i))}
-\tau_{\CFD}\right\}+o_p(1),
\end{split}
\end{equation*}
where $o_p(1)$ is a residual term that converges in probability to zero. We then obtain $\varphi_i^{\wt}$ as the individual summand in the bracket \citep{Tsiatis2006}. A similar reasoning is used to obtain $\varphi^{\text{dr}}_i$ in Proposition 2. Notice that
\begin{equation*}
\begin{split}
\varphi_i^{\text{wt}}+\varphi_i^{\text{dr}}
&=\frac{1}{\pi}\frac{G_i-e(\bfX_i)}{1-e(\bfX_i)}\{2(Y_{i,t+1}-Y_{it})-(\nu(\bfX_i)-\mu(\bfX_i))\}-2\tau_{\CFD},\\
\varphi_i^{\text{wt}}-\varphi_i^{\text{dr}}&=
\frac{1}{\pi}\frac{G_i-e(\bfX_i)}{1-e(\bfX_i)}\{\nu(\bfX_i)-\mu(\bfX_i)\},
\end{split}
\end{equation*}
and the difference in asymptotic variance
\begin{equation*}
\begin{split}
&\mathbb{V}(\varphi_i^{\text{wt}})-\mathbb{V}(\varphi_i^{\text{dr}})
=\E[(\varphi_i^{\text{wt}}+\varphi_i^{\text{dr}}
)(\varphi_i^{\text{wt}}-\varphi_i^{\text{dr}})]
\\
&=\frac{2}{\pi^2}\E\Big[\Big(\frac{G_i-e(\bfX_i)}{1-e(\bfX_i)}\Big)^2(Y_{i,t+1}-Y_{it})(\nu(\bfX_i)-\mu(\bfX_i))\Big]
-\frac{1}{\pi^2}\E\Big[\Big(\frac{G_i-e(\bfX_i)}{1-e(\bfX_i)}\Big)^2(\nu(\bfX_i)-\mu(\bfX_i))^2
\Big],
\end{split}
\end{equation*}
since
\begin{equation*}
\begin{split}
&\frac{2\tau_{\CFD}}{\pi}\E\Big[\Big(\frac{G_i-e(\bfX_i)}{1-e(\bfX_i)}\Big)\Big(\nu(\bfX_i)-\mu(\bfX_i)\Big)\Big]
=\frac{2\tau_{\CFD}}{\pi}\E\Big[\Big(\frac{\E(G_i|\bfX_i)_i-e(\bfX_i)}{1-e(\bfX_i)}\Big)\Big(\nu(\bfX_i)-\mu(\bfX_i)\Big)\Big|\bfX_i \Big]=0.
\end{split}
\end{equation*}
Unfortunately, the expression for $\mathbb{V}(\varphi_i^{\text{wt}})-\mathbb{V}(\varphi_i^{\text{dr}})$ does not further simplify to more elegant forms. But it is evident that there is no guarantee that this difference is nonnegative since it could not be simplified to the expectation of a quadratic form (this is in sharp contrast to the analogous results developed for estimating the average treatment effect, or ATE). Hence even if all models are correct, the double-robust estimator is not necessarily asymptotically more efficient than the weighting estimator.

\bibliographystyle{jasa3}
\bibliography{DR_DID}

\end{document}